\documentclass[reprint,amsmath,amssymb,aps,prl,longbibliography]{revtex4-2}

\usepackage[utf8]{inputenc}
\usepackage[T1]{fontenc}
\usepackage{lmodern}
\usepackage[english]{babel}
\usepackage[autostyle=true]{csquotes}
\usepackage[colorlinks=true,  urlcolor=blue,  linkcolor=blue,  citecolor=blue]{hyperref}
\usepackage[dvipsnames]{xcolor}
\usepackage{siunitx}
\usepackage{graphicx}
\usepackage{txfonts}

\newcommand{\mum}{\rm \si{\micro\meter}}
\newcommand{\ket}[1] {\left\vert #1 \right\rangle}
\DeclareSIUnit\gauss{G}

\newcommand{\aN}{\tilde{N}}
\newcommand{\aP}{\tilde{P}}
\newcommand{\aE}{\tilde{E}}
\DeclareSIUnit\gauss{G}

\begin{document}

\title{Bloch Oscillations of a Soliton in a 1D Quantum Fluid} 

\author{F. Rabec$^{\ddagger}$, G. Chauveau$^{\ddagger}$, G. Brochier, S. Nascimbene, J. Dalibard, J. Beugnon}

\email{beugnon@lkb.ens.fr}

\affiliation{Laboratoire Kastler Brossel,  Coll\`ege de France, CNRS, ENS-PSL University, Sorbonne Universit\'e, 11 Place Marcelin Berthelot, 75005 Paris, France}

\date{\today}

\begin{abstract}
The motion of a quantum  system subjected to an external force often defeats our classical intuition. A celebrated example is the dynamics of a single particle in a periodic potential, which undergoes Bloch oscillations under the action of a constant force\,\cite{Bloch29,Zener34}. Surprisingly, Bloch-like oscillations can also occur in one-dimensional quantum fluids without requiring the presence of a lattice \cite{Gangardt09,Meinert17}. The intriguing generalization of Bloch oscillations to a weakly-bounded ensemble of interacting particles has been so far limited  to the experimental study of the two-particle case, where the observed period is halved compared to the single-particle case\,\cite{Corrielli13,Preiss15}. In this work, we observe the oscillations of the position of a mesoscopic solitonic wave packet, consisting of approximately 1000 atoms in a one-dimensional Bose gas when subjected to a constant uniform force and in the absence of a lattice potential. The oscillation period scales inversely with the atom number, thus revealing its collective nature.  We demonstrate the pivotal role of the phase coherence of the quantum bath in which the wave packet moves and investigate the underlying topology of the associated superfluid currents. Our measurements highlight the periodicity of the dispersion relation of collective excitations in one-dimensional quantum systems. We anticipate that our observation of such a macroscopic quantum phenomenon will inspire further studies on the crossover between classical and quantum laws of motion, such as exploring the role of dissipation, similarly to the textbook case of macroscopic quantum tunneling in Josephson physics \cite{Caldeira81}.
\end{abstract}

\maketitle

\vskip5pt

A single isolated particle moving in a lattice with spatial period $a$ and subjected to a constant force $f$ exhibits Bloch oscillations (BOs) with a period 
\begin{equation}
T_B=\frac{h}{af},
\label{eq:Tbloch}
\end{equation}
where $h$ is Planck's constant. This peculiar motion is due to the periodicity of the lattice band structure. The dynamics in momentum space corresponds to periodic Bragg reflections of the particle when its wavelength matches the period of the lattice. The observation of BOs requires phase coherence during the oscillations and a force weak enough to ensure the adiabatic following of the lowest energy band. It has been observed in ultracold atom platforms  \,\cite{Dahan96,Wilkinson96,Geiger18} as well as in  semiconductor structures with large lattice periods\,\cite{Feldmann92}. 

Bloch oscillations at the single-particle level are extensively used in cold atom experiments for precision measurements\,\cite{Ferrari06,Fixler07,Rosi14,Xu19,Parker18,Morel20} and for characterising the topological properties of band structures\,\cite{Price12,Aidelsburger15}. Classical analogues of  BOs  have also been realized with optical\,\cite{Pertsch99,Morandotti99} and acoustic waves\,\cite{Sanchis07} and in plasmonic waveguide arrays\,\cite{Block14}.

The generalisation of BOs to an ensemble of interacting particles remains largely unexplored. Interactions are often the source of unwanted damping of the oscillations\,\cite{Morsch01,Gustavsson08,Eckstein14,Meinert14}. Nonetheless, the dynamics of repulsively-bound dimers has been explored in Ref.\,\cite{Preiss15} showing effective BOs with half the single-particle period. Similar results have been obtained in a simulation with photons in a waveguide array\,\cite{Corrielli13}. The role of quantum anyonic statistics was  also recently explored in a simulation with an electrical circuit\,\cite{Zhang22}. 

Impurities in one-dimensional (1D) quantum systems \textit{without} a lattice potential also exhibit collective excitations with a periodic behaviour of the energy with momentum\,\cite{Haldane81,Gangardt09,Schecter12,Grusdt14,Will23}, opening the possibility to observe BOs in the absence of a lattice. However, the observation of BOs in bulk 1D systems is challenging due to the dissipation induced by friction terms away from the integrable case\,\cite{Gangardt09}. Recently, an experiment monitoring the dynamics of impurities in a strongly interacting Bose gas showed an evidence of these oscillations\,\cite{Meinert17}.

\vskip2pt
In this work, we study BOs of a large number of interacting particles in a 1D system without any lattice potential. We use a weakly interacting quasi-one-dimensional Bose gas with two components, called $\ket{1}$ and $\ket{2}$. The mesoscopic system is a localized wave packet in component $\ket{2}$ immersed in an extended quantum bath of atoms in  component $\ket{1}$, realizing a so-called magnetic soliton\,\cite{Kosevich90,Qu16,Congy16,Chai20,Farolfi20,Chai22}.  We report the observation of several Bloch oscillations of the position of this two-component soliton. The period of the oscillations is 
\begin{equation}
T=\frac{n_0h}{N_2 f}=\frac{n_0h}{F},
\label{eq:Tsoliton}
\end{equation}
where $n_0$ is the bath density far from the soliton center, $f$ the force acting on a single particle  and $N_2$ is the number of particles in the wave packet. This formula is  reminiscent of  \eqref{eq:Tbloch} but with two major differences: (i) in our case, $T$ depends on the total force $F=N_2f$ acting on the atoms of the wave packet and (ii) the lattice period $a$ is replaced by the inverse of the linear bath density $n_0$. We observe BOs both in a linear and in a ring geometry with uniform total density. In each case, we use matter-wave interference to show that the soliton dynamics are closely related to the phase profile of the bath. In the ring geometry, we explore the interplay between the soliton motion and the pumping of topological supercurrents.

\begin{figure*}[ht!]
\includegraphics[width=0.9\textwidth]{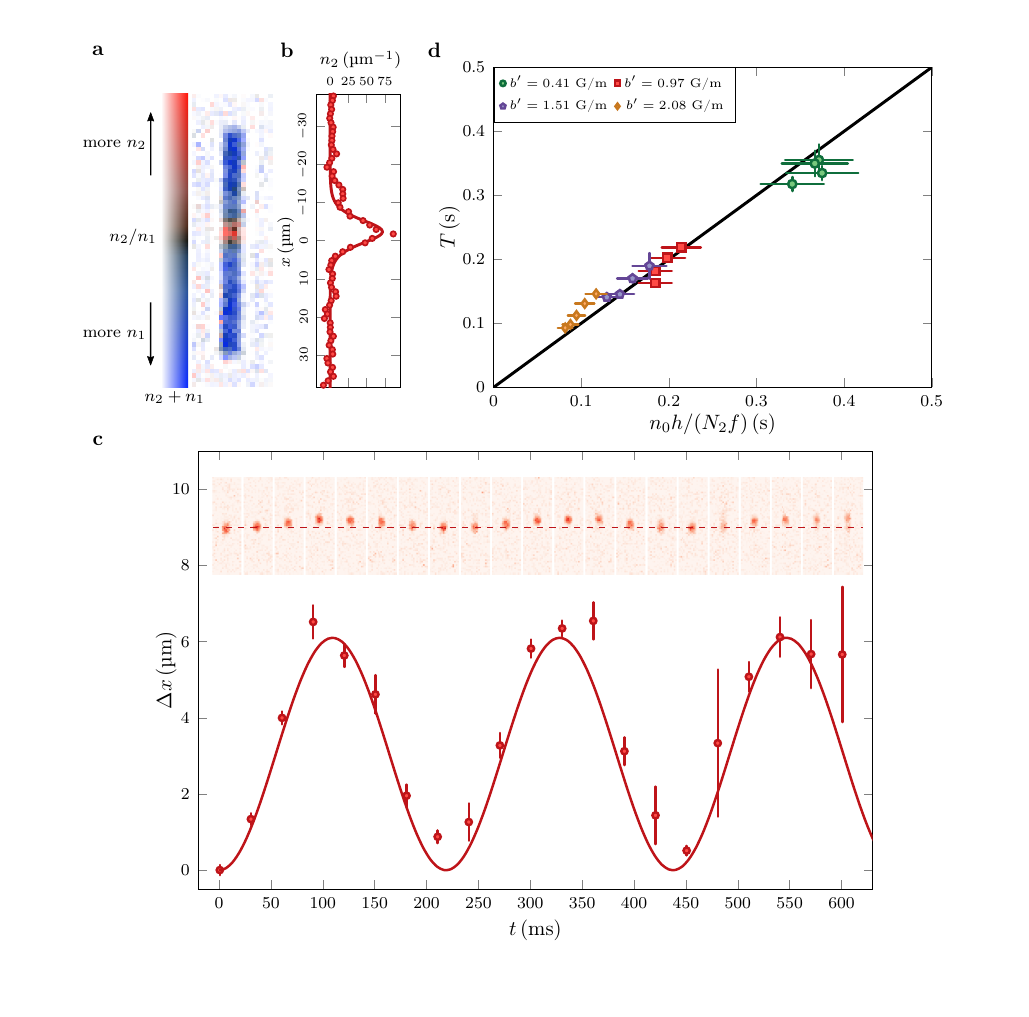}
\caption{\textbf{Soliton preparation and Bloch oscillations in a tube}. \textbf{a}, Reconstructed absorption image of the initial density profiles in both $\ket{1}$ and $\ket{2}$. The colours represent the relative weight of the atomic densities $n_1$ and $n_2$ of both components.   \textbf{b}, Integrated density profile of  component $\ket{2}$ along the transverse direction of the absorption image. The solid line is a fit of the data to a $1/\cosh^2$ function, which corresponds to the expected shape of the soliton in the limit of low values of $n_2/n_1$ at the soliton position. \textbf{c}, Time evolution of the minority component in $\ket{2}$ for $N_2 = 1300(100)$ atoms and $n_0\approx350\,\mum^{-1}$ and under the action of a uniform differential force of $f=6.6(5)\times 10^{-4}\,Ma_\mathcal{G}$, showing the phenomenon of Bloch oscillations. The error bars correspond to the statistical deviation obtained from 10 repetitions of each experiment. The solid line is a sinusoidal fit to the data. Images are shown every 30\,ms from the initial preparation of the wave packet. The dashed line is a guide to the eye to mark the initial position of the wave packet. \textbf{d}, Measured Bloch period when vayring $N_2$ and $f=\mu_B b'$, where $b'$ is the applied magnetic field gradient and $\mu_B$ the Bohr magneton. The different colors correspond to different forces applied on the soliton. The bath density $n_0$ is fixed. The solid black line is the prediction of Eq.\,\eqref{eq:Tsoliton} without any free parameter.}
\label{fig1}
\end{figure*}

\section{Realization of a magnetic soliton}
Our  system consists of a two-component Bose gas of $^{87}$Rb atoms with mass $M$. In the zero temperature limit studied here and for weak interactions, the dynamics of the system is described by two  nonlinear Schrödinger equations (NLSE) coupling the many-body wave functions of each component. The non-linearity is related to the interparticle interaction described by a single intraspecies interaction parameter $g$ for the case studied here and the interspecies interaction parameter $g_i$. We work close to the SU(2) symmetry point for the interactions, corresponding to $g_i\approx g$, and we introduce the spin interaction parameter $g_s=g_i-g\ll g$. The chosen states verify $g_s>0$, meaning that a mixture of these states is weakly immiscible. In the regime $g_s\ll g$, there is a decoupling between the high-energy density modes and  the low-energy spin modes. In the low-energy sector considered here, the total density remains uniform to a good approximation and we focus only on the spin dynamics. The corresponding length scale is $\xi_s=\hbar/\sqrt{2 M g_s n_0}\simeq 2\,\mum$.
The dynamics along the vertical direction is frozen thanks to a strong optical confinement. The in-plane optical confinement is shaped by a spatial light modulator allowing us to design arbitrary trapping geometries. We study quasi-1D systems, either straight lines or rings, of width $d\approx3\,\mum\sim \xi_s$. The typical linear density is $n_0\approx350\,\mum^{-1}$.

We first prepare a gas in the electronic ground state $\ket{1}\equiv \ket{F=1,m=-1}$ at equilibrium in a line of length $L=60\,\mum$. Then, we use a two-photon optical Raman process to transfer part of this gas to  $\ket{2}\equiv \ket{F=1,m=1}$, in a spatially-dependent manner. This  allows us to prepare an arbitrary wave packet of atoms in $\ket{2}$ immersed in a bath of atoms in $\ket{1}$\,\cite{Zou21}, as displayed in  Fig.\,\ref{fig1}ab. By adjusting the initial density profile in $\ket{2}$ and $N_2$, we are able to produce and observe a stationary wave packet with a typical width of $\sigma\approx 5\,\mum$ (see Methods), proving the deterministic realisation of a  magnetic soliton at rest for an immiscible mixture.

\section{Bloch oscillations in a linear geometry}
We then study the response of the magnetic soliton to a state-selective constant force $f$, focusing first on the center of mass dynamics of the wave packet. We report the  observation of BOs in  Fig.\,\ref{fig1}c for a wave packet of $1300(100)$ atoms subjected to a force of $6.6(5)\times10^{-4}\,Ma_\mathcal{G}$, where $a_\mathcal{G}$ is the acceleration of gravity. The measured period $T= 219(3)\,$ms agrees well with the expected value of $210(16)\,$ms. Similar BOs are observed for different values of $N_2$ and $f$ and the measured period shown in Fig.\,\ref{fig1}d agrees well with the predicted $1/(N_2f)$ scaling of Eq.\,\eqref{eq:Tsoliton} without any fitting parameter. It is interesting to emphasize that the period $T$ is independent of the interaction parameters $g$ and $g_i$, thus making it a robust observable. The choice of a weak enough force is necessary to ensure an adiabatic motion of the wave packet,  which should remain close to the family of non-zero velocity solitonic states at all times. This allows us to consider the dynamics of the wave packet as that of a single macroscopic object.  A condition for this adiabatic following is to ensure that the work done by the force across the wave packet extension is  much smaller than the spin interaction energy scale, i.e. $\eta=f\sigma/(g_sn_0)\ll 1$ (here $\eta \sim 0.1$).

%
\begin{figure*}[ht!]
\includegraphics[width=0.8\textwidth]{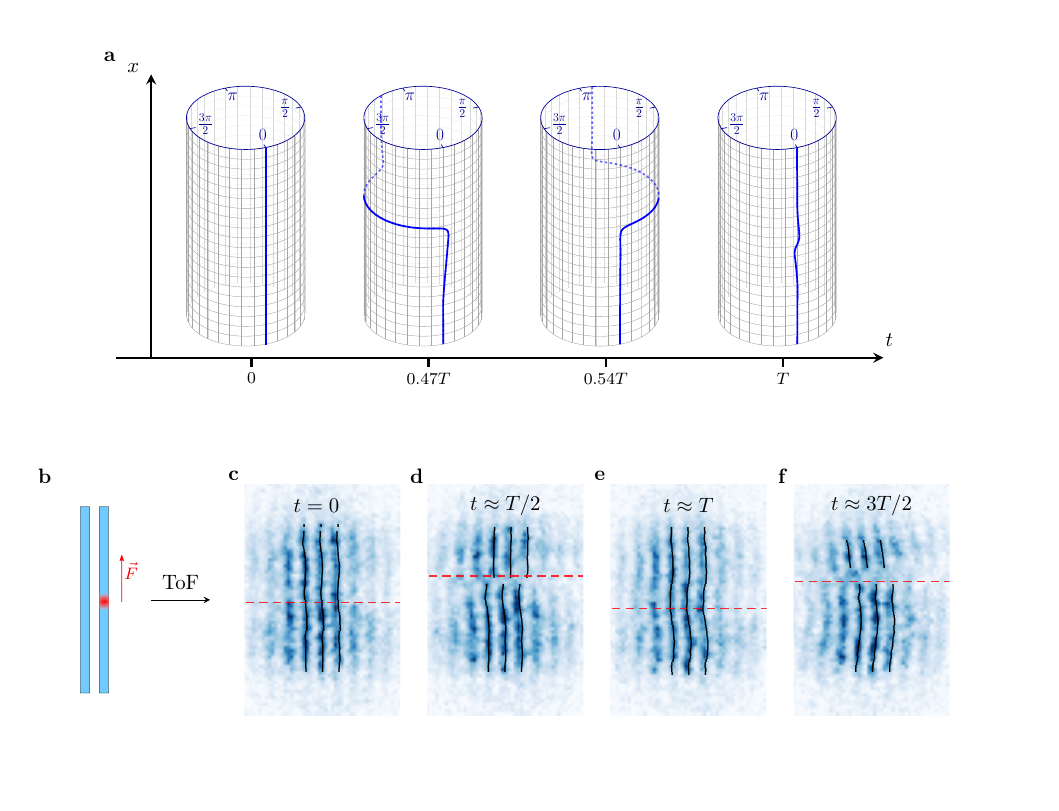}
\caption{\textbf{Evolution of the phase of the bath component during Bloch oscillations.}  \textbf{a}, Calculated phase time evolution corresponding to the experimental case. The line is represented by a dash when it is on the rear surface of the cylinder. At $t=0$, the phase is uniform. At later times, the phase varies on a short distance scale across the wave packet position, where the bath density is the lowest. The phases on both sides of the wave packet are approximately constant and the phase does not wind around the cylinder. The phase is uniform again at $t=T$, giving rise to periodic oscillations. The motion of the wave packet occurs on a short length scale and is thus not visible on the graph.  \textbf{b}, Schematic of the configuration used to perform matter wave interference experiments. The left tube is used as a phase reference with all atoms in $\ket{1}$. The right tube is identical to the left one except for the presence of a localised wave packet in $\ket{2}$.  \textbf{c,d,e,f}, Experimental absorption images of atoms in $\ket{1}$ obtained after a time-of-flight (ToF) expansion from the configuration shown in \textbf{b}. Images are on purpose saturated to highlight the position of the fringes and the color scale is thus qualitative. The matter-wave interference fringes reveal the relative phase between the two clouds. The black lines are the reconstructed positions of the bright fringes. The red dashed lines show the position of the soliton measured independently. A discontinuity of the fringes, corresponding to a $\pi$-phase shift of the phase of the bath is observed at the positions where the wave packet's velocity changes from positive to negative values (\textbf{d,f}). }
\label{fig2}
\end{figure*}

The soliton dynamics can be described with an analytical particle-like model deduced from the NLSE for $g_s\ll g$ \cite{Kosevich01,Congy16,Bresolin23}. We propose here a simple interpretation of the BOs in this limit. The phase of the bath is quasi-uniform  on each side of the soliton or, equivalently, there is no current flowing in the system outside the region of the wave packet. Such a quasi-static equilibrium is reached thanks to the high-energy density modes, which are associated to the propagation of sound at a velocity $c=\sqrt{gn_0/M}$ and thus to a time scale  $L/c\sim20\,$ms, smaller than the Bloch oscillation period. In the limit in which the soliton is not broken apart by the applied force, we introduce the effective spatially-dependent force $f_{\rm eff}(x)$ defined as $ n_0 f_{\rm eff}(x)\equiv n_2(x) f$, where $n_2$ is the density of particles in $\ket{2}$. This leads to the effective potential $ V_{\rm eff}(x)=-\int_{-\infty}^x\mathrm{d}x'\,f_{\rm eff}(x')$ for the system, which varies over the soliton extent only. Introducing the phase of the bath $\Phi_1(x)$, we relate the potential drop between the two edges $x_L$ and $x_R$ of the soliton to the phase difference $\Delta\Phi_1=\Phi_1(x_L)-\Phi_1(x_R)$:
\begin{equation}
\Delta\Phi_1(t)=\frac{1}{\hbar}[V_{\rm eff}(x_L)-V_{\rm eff}(x_R)]t=\frac{Ft}{\hbar n_0}.
\label{eq:Joseph}
\end{equation}
One can show that the phase variation predicted by the particle-like model, in the linear geometry,  is linked to the velocity $V$ of the wave packet by
\begin{equation}
V=V_0 \sin(\Delta \Phi_1),
\label{eq:Vbox}
\end{equation}
in agreement with the periodicity introduced in Eq.\,\eqref{eq:Tsoliton} and with  $V_0=2c_s/\sinh(N_2/n_0 \xi_s)$, where $c_s=\sqrt{g_s n_0/(2M)}$ (see Methods). 

The linear increase of the quantity $P=\hbar n_0 \Delta \Phi_1$with time under the action of the force $F$ shows that it plays the role of a (quasi-)momentum. The expression \eqref{eq:Vbox} for  the velocity  and the one for the  energy $E$  are periodic functions of $P$, as in the case of Bloch oscillations for a particle in a lattice. Such evolution is also analogous to the dynamics of an AC Josephson junction, where the soliton plays the role of a mobile tunnel barrier between two superfluids and the force drives a potential difference between them \cite{Bresolin23}.

The typical evolution of the phase obtained from simulations of the NLSE confirms this picture. We show in Fig.\,\ref{fig2}a the evolution of the phase represented on the surface of a cylinder. Its value is given by the polar angle  and the axis of the cylinder corresponds to the spatial $x$ coordinate of the system.  At time $t=0$, the phase is spatially uniform. At later times, the phase is close to uniform on each side of the wave packet and a phase difference $\Delta \Phi_1$ appears, associated to a strong variation of $\Phi_1(x)$ across the soliton extension. After one period, the phase is approximately uniform again as expected.

We  experimentally confirm the role of the bath phase $\Phi_1$ using matter wave interference (see Fig.\,\ref{fig2}b-f). We prepare a second reference cloud shifted by a few microns with respect to the cloud of interest. This cloud is identical to the first one but it contains only atoms in $\ket{1}$ and constitutes a suitable uniform phase reference for the low-temperature regime studied here. After releasing the two clouds from the trap and allowing them to expand and overlap, we obtain matter-wave interference images. At time $t\approx 0$, we expect both clouds to have a uniform phase, which is confirmed by the observation of straight and parallel fringes. At times $t\approx T/2$ and  $t\approx 3T/2$, on the opposite, we observe a contrasted density hole in the vicinity of the wave packet position associated with a discontinuity in the fringes. This is consistent with the expected full cancellation of the density of the bath at this position and the $\pi$ phase jump at these times.

\begin{figure*}[ht!!]
\includegraphics[width=0.8\textwidth]{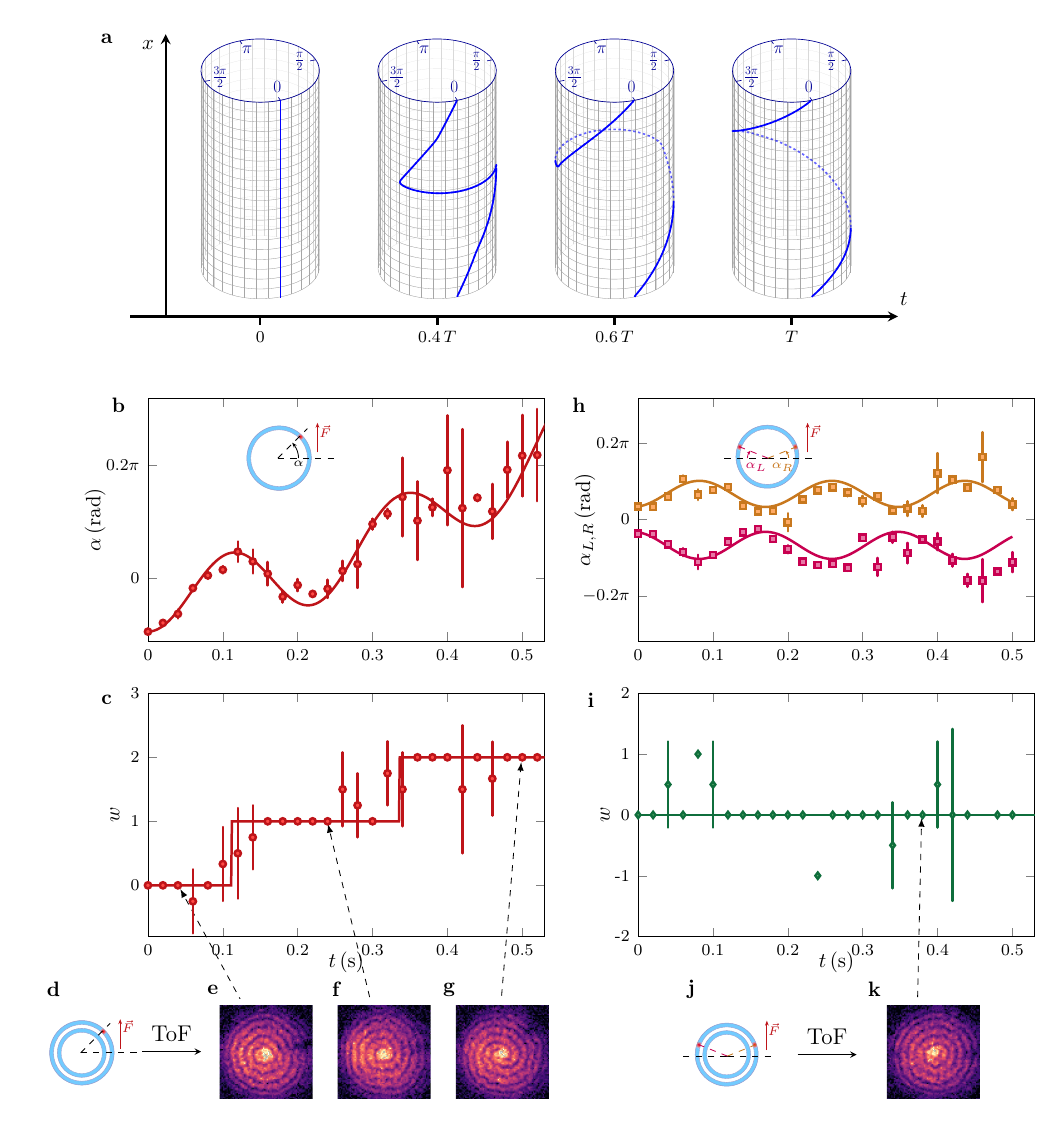}
\caption{\textbf{Oscillation of one and two solitons under a constant force in a ring geometry.} \textbf{a}, Simulated time evolution of the bath phase. Periodic boundary conditions impose the same phase on both sides of the cylinder. Just after $t=T/2$, the phase profile winds once around the cylinder leading to the creation of a supercurrent in the bath. \textbf{b}, Time evolution of the polar angle $\alpha$ associated with the position of the soliton along the ring (see the sketch in the inset) for $N_2=1100(60)$, $n_0=320(10)\,\mum^{-1}$ and $F=6.6(5)\times 10^{-4}\,Ma_\mathcal{G}$. The soliton performs Bloch-like oscillations combined with a drift of the center of mass position due to the generation of supercurrents in the ring. The solid line is the prediction from the particle-like model (see Methods). \textbf{c}, Time evolution of the winding number $w$ of the bath phase. The winding number is measured using matter-wave interference fringes obtained after an expansion from the configuration sketched in \textbf{d} with two concentric rings. Absorption images of the fringes at three different times are shown in  \textbf{e,f,g}. We observe concentric rings (\textbf{e}), an anticlockwise spiral pattern  (\textbf{f}) and an anticlockwise double spiral pattern ( \textbf{g}), which we assign to $w=0$, $w=1$ and $w=2$, respectively.  \textbf{h} Same as in  \textbf{b} but for two solitons initially at diametrically opposed positions denoted by $\alpha_L$ and $\alpha_R$. The two solitons have similar initial atom numbers ($\approx 1500$ atoms). They perform in-phase Bloch oscillations with no clear drift of their positions. The solid lines represent a sinusoidal fit to the data with opposite amplitude.  \textbf{i}, Winding number measured in the two-soliton case. The data are consistent with no observed winding. \textbf{j}, Sketch of the geometry used for matter-wave interference experiments in the two-soliton case. \textbf{k}, Example of measured fringes, corresponding to $w=0$. In all plots, the error bars represent the statistical errors obtained from the 4 repetitions of each experiment.}
\label{fig3}
\end{figure*}

\section{Dynamics in a ring geometry}
The motion of a magnetic soliton under a constant force becomes richer when considering the dynamics in a ring of length $L=150\,\mum$, which corresponds to a situation with periodic boundary conditions. The wave function of the bath being single-valued, this constrains its phase profile. The strong spatial variation $\Delta \Phi_1$ of the phase of the bath across the wave packet position is compensated by a slow linear variation over the rest of the ring. It leads to a so-called backflow current, which carries the total momentum $P=\hbar n_0 \Delta \Phi_1$ of the system  \cite{BecBook2016}. This global current was not allowed in the previous case of the linear geometry with hard walls. Importantly, because of the presence of this backflow current, the properties of the wave packet are no longer invariant under the change $\Delta \Phi_1 \rightarrow \Delta \Phi_1+2\pi$. For example, the velocity of the wave packet, in the particle-like model, now reads
\begin{equation}
V=V_0 \sin\left(\frac{P}{\hbar n_0}\right)+\frac{P}{ n_0 L M},
\label{eq:Vring}
\end{equation}
The second term, $\propto P$, reflects the drift of the wave packet induced by the backflow current. The adiabatic dynamics under the application of a constant force thus mainly consists in an interplay between an oscillatory behaviour and the drag effect induced by the backflow current.

The typical evolution of  the phase profile obtained from simulations of the NLSE is shown in Fig.\,\ref{fig3}a. Periodic boundary conditions impose the phase profile to have the same values on both sides of the cylinder. The backflow appears as a slow winding around the cylinder. Interestingly, after one half-period the profile winds once around the ring. This corresponds to the creation of a topological superfluid current around the ring, induced by the motion of the wave packet. Our system thus realises a new example of a vortex pump \,\cite{Tomoya00,Leanhard02,Wright13}.

To reveal experimentally the role of the backflow momentum and its topological properties, we consider first the dynamics of a single soliton subjected to a uniform force oriented in the plane of the ring.  The projection of the force along the direction of the motion reads $f_\alpha=f\cos \alpha$, where $\alpha$ is the angle describing the position of the wave packet. We restrict here ourselves to situations where $\alpha\ll\pi/2$ and the angular dependence of $f_\alpha$ plays a minor role (Fig.\,\ref{fig3}bc). We measure both the position $\alpha$ (Fig.\,\ref{fig3}b) of the wave packet and the phase winding number $w=\int_0^{2\pi}\mathrm{d}\alpha \,\partial_\alpha \Phi_1(\alpha)/(2\pi)\,\in\mathbb Z$ of the bath (Fig.\,\ref{fig3}c) using matter wave interference with a reference ring (Fig.\,\ref{fig3}d-g)\,\cite{Corman14,Eckel14b}.
As expected, the wave packet exhibits an oscillatory motion, accompanied by a drift in its position induced by the backflow. Each time the wave packet's velocity becomes negative, we observe a jump of one unit in the phase winding of the bath. 
The solid lines on both $\alpha(t)$ and $w(t)$ are obtained from the equations of motions of the particle-like model deduced from Eq.\,\eqref{eq:Vring} and Newton's law, $\mathrm{d}P/\mathrm{d}t=N_2f_\alpha$. They are in excellent agreement with our data.  In the Methods section, we extend this study to the long-time dynamics, for which the angular dependence of $f_\alpha$ plays a significant role. Bloch-like oscillations associated with alternating cycles of pumping and depumping of supercurrents are observed and are superimposed to  a slow periodic and classical-like motion of a pendulum oscillating between $\alpha=0$ and $\alpha=\pi$.

To further investigate the soliton dynamics in a ring, we performed an experiment with two identical solitons, initially positioned diametrically opposed to one another. Their positions are defined by the angles $\alpha_L$ and $\alpha_R$ with respect to their initial positions. As illustrated in  Fig.\,\ref{fig3}h, the solitons exhibit synchronized oscillations as expected from the relative orientation of the force applied to them. No drift in the solitons' positions is observed in this case. This is consistent with the measurement of the bath phase winding (Fig.\,\ref{fig3}i-k) which shows no significant creation of superfluid currents. This behaviour reflects that the current generated by each soliton motion are opposite in sign and thus cancel each other out.

\section{Discussion and outlook}
In this work, we have observed Bloch oscillations of a magnetic soliton and demonstrated the crucial role of the phase of the bath in the dynamics of the system. This phase can be associated to a backflow current induced by the motion of the wave packet itself. When multiple solitons are generated in the ring, this current leads to mediated interactions between the solitons through the bath.  In the context studied here, where the fast density modes are decoupled from the dynamics studied in the spin sector, these interactions are quasi-instantaneous and non-local.  Besides the particular case explored experimentally in this work,  where the backflow is cancelled (Fig.\,\ref{fig3}h-i), the total energy of a set of identical solitons will include a $\sum_{i,j} P_i P_j$ term. This Ising-like term occurs in many other situations in classical physics (chemistry, biology, neural networks, chaotic systems), although it usually involves local variables like (pseudo-)spins instead of the momenta of the particles\,\cite{Kuramoto02}. For interacting magnetic solitons, it can lead  to rich non-linear dynamics at the frontier between classical and quantum dynamics that could be explored in future work. It also opens new avenues to the emerging field of atomtronics, where the role of solitons, supercurrents and gauge fields plays a central role \cite{Amico22} and it provides new perspectives to develop strong analogies between atomic systems and superconducting devices in condensed matter settings.

\vskip10pt
\begin{acknowledgments}
\textit{Acknowledgments.} 
G.C. and F.R. contributed equally to this work. We acknowledge the support by ERC (Grant Agreement No 863880). We thank N. Pavloff for sharing personal notes and N. Cooper and M. Fleischhauer for fruitful discussions and C. Heintze for his participation at the early stage of the project.
\end{acknowledgments}

\definecolor{violet3}{RGB}{120,0,180}
\hypersetup{urlcolor=violet3} 

\bibliography{bibliography_magneticsoliton}


\section{METHODS}

\subsection{Experimental details}

\paragraph{Preparation of the linear sample.}
We prepare a single tube of  $N_1\sim2\times 10^4$ $^{87}$Rb atoms of mass $M$ in an optical dipole trap. The tube length is $L=60\,\mum$ and its linear density $n_0=N_1/L$ is uniform along the axis $x$ of the tube. This gas is obtained starting from an horizontal uniform planar Bose gas\,\cite{Ville17}. The dynamics along the vertical direction $z$ is frozen in an approximate harmonic potential of frequency $\omega_z=2\pi \times 4.3(1)\,$ kHz, which is created by an optical lattice corresponding to a characteristic size of the ground state wave function $\ell_z=\sqrt{\hbar/M\omega_z}\simeq	160\,$nm.  In the atomic plane, the gas is initially  confined in an optical box potential whose rectangular shape is controlled thanks to a digital micromirror device (DMD), see Refs.\,\cite{Ville17,Bakkali21} for details. We then slowly modify the shape of the box potential to reach the linear geometry by displaying on the DMD a series of  50 images every 10 ms that interpolates between the initial and final configurations. The final transverse confinement in the horizontal plane, along $y$, is given by the projection of a flat-bottom optical potential of size $\sigma_y=3\, \mum$ programmed on the DMD. In practice, this potential is smoothed by the optical system response which projects the image from the DMD onto the atomic gas with a typical resolution of $1\,$\mum. The temperature of the initial 2D cloud is the lowest achievable in the experiment, typically below $20\,$nK. The gas is in the weakly-interacting regime with a Lieb-Liniger parameter $\gamma=Mg/(n_0\hbar^2)\approx 10^{-4}$\,\footnote{This value of $\gamma$ should be considered with care because the gas is not strictly in the 1D regime.}
, where $g\approx 3\times 10^{-39}$\,J.m is the 1D interaction parameter\,\cite{Lieb63}. 
Atoms are initially prepared in the $\ket{1}=\ket{F=1,m=-1}$ electronic ground state and constitutes the bath of atoms. A magnetic field of amplitude $B_0\approx 2\,$G along $y$ lifts the degeneracy of the Zeeman states. We then prepare a localized wave packet of atoms in the lowest-energy state $\ket{2}=\ket{F=1,m=1}$ thanks to a two-step transfer via $\ket{F=2,m=0}$. The first step is a spatially-resolved two-photon Raman transfer allowing us to choose the spatial profile $n_2(x)$ of the transferred cloud, keeping the total density $n_0=n_1(x)+n_2(x)$ constant \cite{Zou21,Bakkali21}. The second step is performed by a microwave transfer with no spatial resolution and which does not affect the atoms in state $\ket{1}$. By symmetry, $\ket{1}$ and $\ket{2}$ have the same intra-species  scattering length $a_1=(100.4-0.18)\,a_0$, where $a_0$ is the Bohr radius. The interspecies scattering length is $a_i=(101.3+0.18)\, a_0 $ \cite{vanKempen02}. For both scattering lengths, we have added the correction due to magnetic dipole-dipole interactions, which are non-negligible in this geometry because of the strong confinement  along the $z$ direction \cite{Zou20a}. This mixture is thus weakly immiscible ($a_i>a_1$). Its immiscibility is enhanced by magnetic dipole-dipole interactions compared to the case of Ref.\,\cite{De14}, where demixing dynamics have been studied in a 3D geometry. The natural length scale to describe spin dynamics in this mixture is given by the ``spin-healing'' length $\xi_s=\hbar/\sqrt{2 M n_0 |g_s|}\approx 1.7\,\mum$, where $M$ is the atomic mass and $g_s \propto (a_1-a_i)$ is the effective spin-interaction parameter in the quasi-1D geometry studied here ($\xi_s \sim \sigma_y$). Its precise value depends on the detailed shape of the trapping potential along $y$ and its calibration is discussed below.

We apply a constant and uniform force on the system using an external magnetic field with a linear variation along $x$. This magnetic field gradient $b'$, of typical amplitude 1\,G/m, is switched on slowly before the partial transfer of atoms to the minority component. The associated force is opposite for the  states $\ket{1}$ and $\ket{2}$ because of their opposite magnetic moment $\pm \mu_B/2$, with $\mu_B$ the Bohr magneton. The differential force between the two species is thus $f=\mu_B b'\sim 10^{-28}\,$N. The associated energy difference between both ends of the tube is $\sim \mu _B b' L\sim 1\,$nK, which is small compared to all other relevant energy scales in the problem. We transfer part of the atoms from $\ket{1}$ to $\ket{2}$ at time $t=0$. Atoms in state $\ket{2}$ thus experiences an abrupt application of the force. We checked numerically that in the range of parameters explored in this work, the dynamics of the wave packet is given by the difference of force $f$ between the two components.

\paragraph{Calibration of the force.}
We calibrate the magnetic field gradient $b'$ by performing Ramsey spectroscopy. We use a microwave field at a frequency of $\sim6.8\,$GHz, resonant with the $\ket{F=1,m_1=-1}~\rightarrow~\ket{F=2,m_2=-2}$ transition. This transition being magnetic-field-sensitive, the magnetic field gradient causes a spatial dependence of the transition frequency. We measure it performing two identical microwave pulses, separated by a wait time $t_w$. We then image the $\ket{F=1,m=-1}$ component and observe interference fringes with a spatial period $\Lambda$. The magnetic gradient is given by $|b'|=\left|h/[\Lambda t_w\mu_B(g_{1}m_1-g_2m_2)]\right|=\left|2h/(3 \Lambda t_w\mu_B)\right|$, where $h$ is the Planck constant and $g_{1,2}=\pm 1/2$ the Landé factors.

\paragraph{Preparation of a ring trap and measurement of the winding.}
We measure the winding number of the supercurrents in a ring geometry using matter-wave interferometry\,\cite{Corman14,Eckel14b, Aidelsburger17}. We use a two-ring geometry. The inner ring serves as a phase reference where all the atoms are in $\ket{1}$. The soliton is prepared in the outer ring. To create this geometry we start with a $30\,\mum$ disk. This disk is then shrinked dynamically by displaying a movie on the DMD. We obtain a ring of inner radius $16\,\mum$ and outer radius $25\,\mum$. Then, we split this ring into two rings by adding a $3\,\mum$ thick circular barrier with a central radius of $20.5\,\mum$. This procedure is meant to avoid  supercurrents in either of the two rings. We estimate the frequency of unwanted windings around $2\%$.   
To obtain the winding number, we make the reference ring interfere with the other ring: we lower the power of the vertical optical lattice by a factor 3, we then switch off the in-plane confinement during $4.5\,$ms before doing a $1\,$ms time-of-flight in the absence of the residual vertical confinement. We finally perform absorption imaging to observe the interference patterns and determine the winding number $w$ of the spiral patterns.

\subsection{Preparation of a magnetic soliton at rest}
\begin{figure}[t!]
\includegraphics[width=0.49\textwidth]{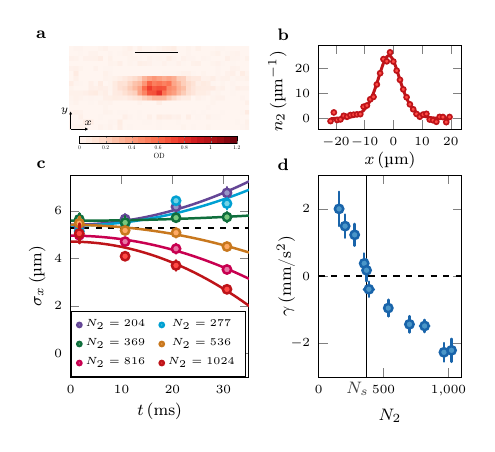}
\caption{\textbf{Magnetic soliton at rest.} \textbf{a}, Absorption image of the minority component wave packet in $|2\rangle$. The axis of the tube is horizontal. The color is in arbitrary units proportional to the atomic surface density. The horizontal solid line corresponds to a length of $10\,\mum$. \textbf{b}, Mean density along the tube direction ($x$). The solid line is a fit of the data to the analytical profile of the magnetic soliton.  \textbf{c}, We vary the atom number transferred  to the minority component and monitor the short time evolution of the size of the wave packet. Solid lines are fits of the function $t \mapsto \sigma_0+\gamma t^2$ to the data. \textbf{d}, Evolution of the expansion coefficient $\gamma$ as a function of the atom number $N_2$. The value $N_s$ corresponds to the atom number for which the wave packet is stationary. This experiment thus demonstrates the realization of a magnetic soliton at rest.}
\label{figs1}
\end{figure}

We demonstrate in this section the experimental realization of a stationary solitonic wave packet and use this measurement to calibrate the value of $ g_s$. For a given size $\sigma$ of the wave packet of atoms in state $\ket{2}$ created with the two-photon transfer, we vary its atom number $N_2$ and study the short time evolution of the size of the wave packet. 
In the limit of low depletion, \textit{i.e.} $N_2 \rightarrow 0$, the stationary profile of the minority component is given by $n_2(x)=N_s/[2\sigma\cosh^2(x/\sigma)]$, where $\sigma=\hbar^2/(m g_s N_s)$. For a given size, if $N_2<N_s$, the kinetic energy of the system is larger than the effective interaction energy and the wave packet expands. For a larger atom number, the effective attractive interaction is larger than the kinetic energy and the wave packet contracts. We confirm this evolution in Fig.\,\ref{figs1}c. For $N_2 \simeq 400$, the size does not evolve significantly over $30$ ms and a stationary magnetic soliton is prepared. More quantitatively, we fit the data to the function $t \mapsto \sigma_0+\gamma t^2$. The fitted expansion coefficients $\gamma$ are shown in Fig.\,\ref{figs1}d. From this measurement we obtain the atom number $N_s\approx370(60)$ corresponding to a stationary profile when $\gamma=0$. Using this value, we determine $ g_s=4.0(6)\times10^{-41}$\,J.m. The obtained value is in good agreement with the expected value for the chosen mixture in this tube geometry.  

\subsection{From NLSE to a particle-like model}
In this section, we derive the particle-like model describing the properties and the dynamics of the magnetic soliton. We restrict ourselves to situations where the system size $L$ verifies $L \ll c t$, where $c$ is the speed of sound associated with the density modes and $t$ is the typical time on which the dynamics of the soliton is studied. This allows us to assume that on the slow time scale of the spin dynamics, the bath, and in particular its phase, has  reached a quasi-equilibrium situation. The case  $L \gtrsim c t$ is discussed in the section dedicated to numerical simulations.

\paragraph{Equation of motion without external force.}
We start from the description of the  two-component system by  coupled NLSE for the matter wave fields $\psi_{1,2}$:
\begin{equation}
	\left\{
	\begin{array}{ll}
		-\frac{\hbar^2}{2 m} \nabla^2 \psi_1 + g \, n_1 \, \psi_1 + g_{i} \, n_2 \, \psi_1 = \mathrm{i} \hbar \, \frac{\partial \psi_1}{\partial t} \\[5pt]
		-\frac{\hbar^2}{2 m} \nabla^2 \psi_2 + g \, n_2 \, \psi_2 + g_{i} \, n_1 \, \psi_2 = \mathrm{i} \hbar \, \frac{\partial \psi_2}{\partial t},
	\end{array}
	\right.
	\label{eq: coupled_GPEs_magnetic}
\end{equation}
where we have introduced the atomic densities $n_i(x,t)=|\psi_i(x,t)|^2$, with $i=\{1,2\}$ and chosen $\int \mathrm{d}x\, |\psi_i(x,t)|^2=N_i$.

We assume that the total density $|\psi_1|^2 + |\psi_2|^2 $ remains constant, a good approximation in the Manakov limit $g_s \ll g$. We parametrize the problem with 3 real variables $\theta$, $\varphi$, $\Phi$:
\begin{equation}
\begin{pmatrix}
	\psi_1 \\
	\psi_2
\end{pmatrix} = \sqrt{n_0} e^{i\Phi/2} \begin{pmatrix}
	\cos(\theta/2)e^{-i\varphi/2}\\
	\sin(\theta/2)e^{+i\varphi/2}\\
\end{pmatrix}.
\end{equation}
The total particle current is defined as
\begin{equation}
\chi(x,t)=\frac{\hbar}{M}\,  \mathcal{I}(\psi_1^\ast \partial_x \psi_1 +  \psi_2^\ast \partial_x \psi_2), 
\end{equation}
 where $\mathcal{I}(z)$ is the imaginary part of $z$. Under the assumption of constant total density, the continuity equation yields $\partial_{ x}\,\chi=0$.  The total particle current is thus space-independent.

We now set the spin healing length $\xi_s=\hbar/\sqrt{2n_0 Mg_s}$ as the length unit  and $\tau_s=\hbar/(n_0g_s)$ as the time unit. In the following, we use the $\tilde \cdot$ notation for dimensionless quantities.
We obtain 
\begin{equation}
\tilde \chi(\tilde x, \tilde t)=n_0 \xi_s\left[\cos^2(\theta/2)\left(\partial_{\tilde x}\Phi-\partial_{\tilde x}\varphi\right) +\sin^2(\theta/2)\left(\partial_{\tilde x}\Phi+\partial_{\tilde x}\varphi\right) \right].
\label{eq:current}
\end{equation}
For convenience, we introduce $\tilde J=\tilde \chi/(n_0 \xi_s)$. Using Eq.\,\eqref{eq:current}, we get the relation
\begin{equation}
\partial_{\tilde x}\Phi = \cos(\theta)\,\partial_{\tilde x} \varphi + \tilde J.
\end{equation}

We can now rewrite the coupled NLSE given in Eq.\,\eqref{eq: coupled_GPEs_magnetic} as a function of the three variables  $\theta$, $\phi$ and $\tilde J$:
\begin{equation}
	\left\{
	\begin{array}{ll}
		\partial_{\tilde t}\, \theta = -2\cos(\theta)(\partial_{\tilde x}\varphi)(\partial_{\tilde x}\theta) - \sin(\theta)\partial_{\tilde x}^2\varphi - \tilde J \partial_{\tilde x}\theta  \\[5pt]
		\partial_{\tilde t}\,\varphi =-\cos(\theta)\left[1+\left(\partial_{\tilde x}\varphi\right)^2\right]+\dfrac{\partial_{\tilde x}^2\theta}{\sin(\theta)}-\tilde J\partial_{\tilde x} \varphi.
	\end{array}
	\right.
	\label{coupled_theta_phi}
\end{equation}

\paragraph{Conserved quantities.}
We define three conserved quantities for the previous system of equations: 
\begin{alignat}{3}
	\aN &=&\, \frac{1}{2}&\int \mathrm{d}\tilde x \left[ 1-\cos(\theta)\right] \\
	\aP &=&\, \frac{1}{2}&\int \mathrm{d}\tilde x \left[ 1-\cos(\theta)\right]\partial_{\tilde x}\varphi \\
	\aE &=&\, \frac{1}{4}&\int \mathrm{d}\tilde x\left\{\left(\partial_{\tilde x} \theta\right)^2 + \sin^2(\theta)\left[1+\left(\partial_{\tilde x}\varphi\right)^2\right] + \tilde J^2 \right\}.
\end{alignat}
The quantity $\tilde N=N_2/(n_0 \xi_s)$ is associated with the atom number in the wave packet.
The quantity $\tilde P$ can also be expressed in terms of the phase of the bath $\Phi_1=(\Phi -\varphi)/2$:
\begin{equation}
\aP=-\int\mathrm{d}\tilde x\, \partial_{\tilde x}\Phi_1 + \int\mathrm{d}\tilde x\, \tilde J/2 \label{eq:PphiJ}
\end{equation}
and we will check that it is the canonical momentum of the system. The quantity $\aE$ corresponds, up to an overall constant term, to the total energy of the system.

\paragraph{Solitonic solutions.} We look at solitonic solutions of Eq.\,\ref{coupled_theta_phi} for which component 2 is localized and that have the form ${\theta(\tilde x,\tilde t)=\theta(\tilde x-\tilde V \tilde t)}$, ${\varphi(\tilde x,\tilde t)=\Omega \tilde t +\bar{\phi}(\tilde x-\tilde V\tilde t)}$, where $\tilde V$ is the velocity of the soliton in the laboratory frame. This family of solutions, parametrized by $(\Omega, \tilde V)$, is usually named magnetic solitons and has been investigated for example in Ref.\,\cite{Kosevich90} in the case $\tilde J=0$.  It can be shown that for an arbitrary value of $\tilde J$ we have:
\begin{align}
	\cos[\theta(\tilde x)]&=1-\frac{4\kappa^2}{2+\Omega + \sqrt{\Omega^2 + u^2}\cosh(2\kappa \tilde  x)} \\ 
	\bar{\phi}(\tilde x)&=\frac{1}{2}u\tilde  x  + \nonumber \\ &\arctan\left[\frac{2\Omega-u^2+2\sqrt{\Omega^2 + u^2}}{2\kappa u}\tanh(\kappa \tilde  x)\right],
\end{align}
where ${u=\tilde V-\tilde J}$ is the velocity of the soliton in the frame moving with the current in the bath and ${\kappa=\sqrt{1+\Omega-u^2/4}}$. The previous solitonic solutions exist provided that ${1+\Omega > u^2/4}$. We notice that the function $\bar{\phi}$ is not defined for $u=0,\Omega>0$. Indeed, we have 
\begin{equation}
\lim_{u\to 0^+, \Omega>0}\,\bar{\phi}=-\lim_{u\to 0^-, \Omega>0}\,\bar{\phi}=-\frac{\pi}{2}+ \pi H(\tilde x),
\end{equation}
where $H$ is the Heaviside function. These two solutions describe the same physical situation.

\begin{figure}
\includegraphics[width=0.49\textwidth]{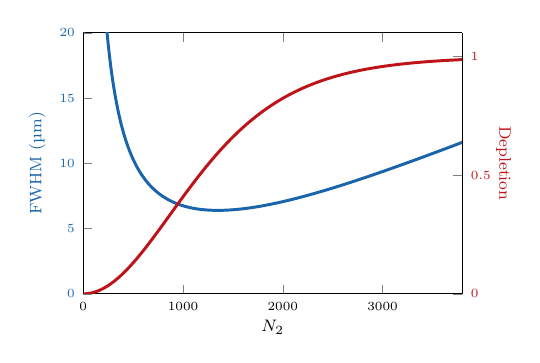}
\caption{\textbf{Full width at half maximum and depletion of a magnetic soliton.} We focus on the soliton at rest ($\tilde v=0$ and $\tilde J=0$). This graph is plotted using typical parameters of the experiment: $n_0 = 330 \, \mum^{-1}$ and $\xi_s = 2 \, \mum$.}
\label{figfwhm}
\end{figure}

We illustrate the properties of the soliton at rest by showing in Fig.\,\ref{figfwhm} the evolution of the full width at half maximum (FWHM) size of the wave packet in $\ket{2}$ as a function of $N_2$ for our typical experimental parameters. The associated depletion of the bath, corresponding to $1- \mathrm{min}_x \, [n_1(x)/n_0]$ is also plotted on the same graph. 

The expression of the conserved quantities $\aN, \aP$ and $\aE$ in terms of the parameters $\Omega$ and $u$ are given by 
\begin{align}
	\aN&=\ln\left( \frac{2+\Omega+2\kappa}{\sqrt{\Omega^2+u^2}} \right) ; \label{eq:N}\\
	\aP&=2\arctan\left(\frac{\kappa u}{\sqrt{\Omega^2 + u^2}-\Omega +\frac{u^2}{2}}\right) ;\label{eq:P}\\
	\aE&=2 \kappa + \int\mathrm{d}\tilde x\, \tilde J\,^2. \label{eq:E}
\end{align}
Combining Eqs.\eqref{eq:N}, \eqref{eq:P} and \eqref{eq:E}, we obtain the dispersion relation of the magnetic soliton:
\begin{equation}
\aE(\aN, \aP) = 2\tanh(\aN/2) + 4\frac{\sin^2(\aP/2)}{\sinh(\aN)} +\int\mathrm{d}\tilde x\,\, (\tilde J\,/2)^2.
\label{eq:energy_soliton}
\end{equation}
Inverting Eq.\,\eqref{eq:P}, one gets the velocity
\begin{equation}
u=2\frac{\sin \tilde P}{\sinh \tilde N}.\label{eq:veloc}
\end{equation}
This confirms that the quantity $\tilde P$ verifies the definition of a canonical momentum
\begin{equation}
u=\frac{\partial \tilde E}{\partial \tilde P}\bigg|_{\tilde N}.
\end{equation}

We remind that the results derived so far are independent of the geometry (line or ring) and assume only a system of sufficiently short length $L$, for which the particle current $\tilde J$ is spatially uniform. We now consider separately the situations studied in the main text. The linear and ring geometries correspond to strict and periodic boundary conditions, respectively.

\paragraph{Linear geometry.}
We consider in this paragraph a system of length $\tilde{L}$ with strict boundary conditions and we assume that the total current $\tilde J(t)$ vanishes at all times. The quantity $\tilde P$ is given by
\begin{equation}
\tilde P=-\int_0^{\tilde{L}}\mathrm{d}\tilde x \,\partial_{\tilde x}\Phi_1(\tilde x)=\Phi_1(
0)-\Phi_1(\tilde L)=\Delta\Phi_1.
\end{equation}

Moreover, under the assumption of zero total current, the phase of the bath is uniform away from the localized wave packet of the minority component, \textit{i.e} when $\cos(\theta)=0$, as can be seen using the continuity equation. 
The energy of the soliton reads
\begin{equation}
\aE(\aN, \Delta\Phi_1 )= 2\tanh(\aN/2) + 4\frac{\sin^2(\Delta\Phi_1/2)}{\sinh(\aN)}.
\end{equation}
This relation is periodic in $\tilde P=\Delta\Phi_1$ and this property is at the origin of Bloch oscillations.

\paragraph{Ring geometry.}
For a system with periodic boundary conditions, we can parametrize the phase such that $\Phi_1(-\tilde{L}/2)=\Phi_1(\tilde{L}/2)$. Using Eq.\,\eqref{eq:PphiJ} and the fact that $\partial_{\tilde x}\tilde J=0$, the quantity $\aP$ is then given by
\begin{equation}
\aP= \tilde J \tilde{L}/2.
\end{equation}
Inserting this expression in Eq.\,\eqref{eq:energy_soliton}, we obtain
\begin{equation}
\aE(\aN, \aP)= 2\tanh(\aN/2) + 4\frac{\sin^2(\aP/2)}{\sinh(\aN)} + \frac{\aP^2}{\tilde{L}}.
\end{equation}
Due to the presence of the last term, the energy is thus not a periodic function of $\tilde P$.

\paragraph{Action of a force.}

We now consider the case where an additional force acting only on atoms in state $\ket{2}$ is applied.  The following discussion applies both to the linear and the ring geometries. The force is associated with a potential $\tilde{U}(\tilde x)$ and leads to a modification of Eq.\,\eqref{coupled_theta_phi}:
\begin{equation}
\partial_{\tilde t}\varphi =-\cos(\theta)\left[1+\left(\partial_{\tilde x}\varphi\right)^2\right]+\frac{\partial_{\tilde x}^2\theta}{\sin(\theta)}-\tilde J\partial_{\tilde x} \varphi -\tilde U.
\end{equation}
We now make the adiabatic approximation which assumes that the spatial variation of the potential $\tilde U$ is small over the extent of the soliton. Then, the wave packet's profile remains given by a solitonic solution of equation (\ref{coupled_theta_phi}). 
In this case, the quantities $\aE, \aN$ are still conserved quantities but the expression of the energy is modified:
\begin{multline}
E = \frac{1}{4}\int \mathrm{d}\tilde x \left\{\right. \left(\partial_{\tilde x} \theta\right)^2 \\+ \sin^2(\theta)\left[1+\left(\partial_{\tilde x}\varphi\right)^2\right] + \tilde J\,^2 + 2(1-\cos{\theta})\tilde U \left.\right\}.
\end{multline}
The quantity $\aP$ is not conserved anymore and evolves
\begin{equation}
\partial_{\tilde t} \aP = -\frac{1}{2}\int\mathrm{d}\tilde x\left[1-\cos(\theta)\right]\partial_{\tilde x} \tilde{U}. \label{eq:PVeff}
\end{equation}
We again use the adiabatic approximation and neglect the work of the force over the extent of the soliton. This leads to
\begin{equation}
\partial_{\tilde t} \tilde P = -\partial_{\tilde x} \tilde{U}|_{\tilde x=\tilde x_0} \frac{1}{2}\int\mathrm{d}\tilde x \, [1-\cos(\theta)]=-\tilde N\partial_{\tilde x} \tilde{U}|_{\tilde x=\tilde x_0},
\end{equation}
where $\tilde x_0$ is the position of the center of the wave packet. We obtain that the canonical momentum $\tilde P$ obeys Newton law, as expected. The energy of the soliton under the adiabatic approximation is given by 
\begin{equation}
	E( \aN, \aP) = 2\tanh(\aN/2) + 4\frac{\sin^2(\aP/2)}{\sinh(\aN)} + \int\mathrm{d}\tilde x\, \tilde J^2/4 + \aN\tilde{U}(\tilde x).
	\label{energy_soliton}
\end{equation}

\begin{figure*}[t!]
\includegraphics[width=\textwidth]{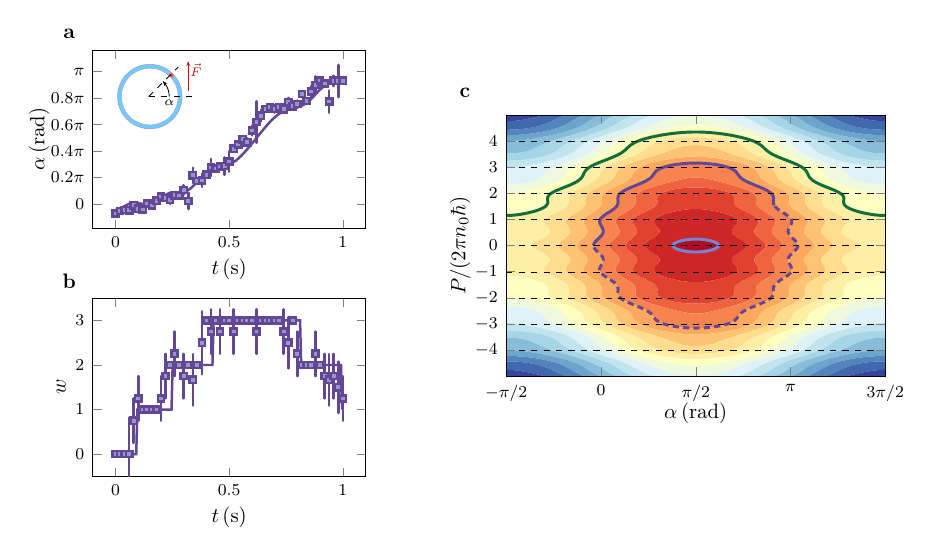}
\caption{\textbf{Long time evolution of a soliton in a ring.} \textbf{a}, Angular position of the wave packet moving from one side of the ring to the other for $N_2=1600(80)$, $n_0=320(10)\,\mum^{-1}$ and $F=6.6(5)\times 10^{-4}\,Ma_\mathcal{G}$.\textbf{b}, Corresponding phase winding of the bath. The winding increases up to +3 and then decreases when $\alpha >\pi/2$. In both plots, the error bars represent the statistical errors obtained from the 4 repetitions of each experiment and the solid line is the prediction of the particle-like model. \textbf{c}, Calculated phase portrait $(\alpha, P)$ of the soliton motion starting at $\alpha=P=0$ (violet line). The evolution is periodic. The first part of the computed motion (solid line part) corresponds to the experimental observation of figures \textbf{a}\textbf{b}. The dashed part of the line shows the expected subsequent evolution. The wave packet reaches the opposite side of the ring ($\alpha=\pi$) with zero momentum and then returns to its initial position with its momentum taking negative values. The two other solid lines correspond to different initial conditions. For $\alpha(t=0)$ close to $\pi/2$ and with zero initial velocity, we obtain a classical-like trajectory (blue ellipse). For a sufficiently large initial velocity, we obtain open orbits (green solid line). The colored background contour plot is associated with different values of the energy of the system with an arbitrary scale. The soliton trajectories are associated with iso-energy curves.}
\label{fig4}
\end{figure*}

In the case $\tilde J=0$ (corresponding to linear geometry of the main text) and a constant and uniform force, we obtain the equations describing Bloch oscillations. We have
\begin{equation}
\frac{\mathrm{d}\tilde P}{\mathrm d \tilde t}=\tilde F,\label{eq:Newton}
\end{equation}
where $\tilde F=\tilde N \tilde f$ and $\tilde f=-\partial_{\tilde x} \tilde{U}$.
For a soliton initially at rest and a uniform force, we obtain the time evolution of the position of the wave packet:
\begin{eqnarray}
\tilde X(\tilde t)=2\frac{1-\cos(\tilde F \tilde t)}{\tilde F \sinh(\tilde N)}
\end{eqnarray}
The dimensionless period is thus given by
\begin{eqnarray}
\tilde T=\frac{2\pi}{\tilde N \tilde f}.
\end{eqnarray}
\vskip5pt

For $\tilde J\neq0$ (corresponding to the ring geometry in the main text), Eq.\,\eqref{eq:Newton} is still valid and Eq.\,\eqref{eq:veloc} is modified into
\begin{eqnarray}
\tilde V= 2 \frac{\sin(\tilde P)}{\sinh(\tilde N)}+2\frac{\tilde P}{\tilde L}.\label{eq:Veloc}
\end{eqnarray}

\paragraph{Expressions in physical units.}
The dimensionless expressions given in the previous paragraph can be  written in physical units, as in the main text, by rescaling distances by $\xi_s$, times by $\tau_s$, velocities by $c_s=\xi_s/\tau_s$. The physical current, energy, force and momentum are given by $\chi=n_0 c_s \tilde J$, $E=n_0\hbar c_s \tilde E$, $f=\hbar^2/(2M \xi_{s}^3)\tilde f$ and $P=n_0 \hbar \tilde P$, respectively.

\subsection{Long-time dynamics in the ring}
In this section we discuss the long-time dynamics of a single soliton in a ring, where the angular dependence of $f_\alpha$ now plays a role. The position of the wave packet and the winding evolution are shown in Fig.\,\ref{fig4}a and Fig.\,\ref{fig4}b, respectively. The time range studied experimentally is limited to $\approx 1\,$s by atom losses and finite temperature effects. Oscillations with a drift are also observed and the winding number increases up to $w=3$. The additional feature with respect to the situation explored in the main text is the decrease of the winding number $w$ when $\alpha>\pi/2$. Indeed, in this range, $f_\alpha<0$ and Bloch-like oscillations induce a depumping of the topological current. These observations are in excellent agreement with the predicted equations of motion of the particle-like model shown as a solid line.

We extend the study of the long time evolution  theoretically using the phase portrait $(\alpha, P)$ shown in Fig.\,\ref{fig4}c. The motion of the wave packet, represented by the violet line, corresponds to a constant energy curve. After the first part of the experimentally observed evolution (solid line), the wave packet reaches $\alpha=\pi$ with $w=0$. The wave packet then returns to $\alpha=0$ (dashed line) with a winding that first decreases to $w=-3$ and then returns to its initial value of $w=0$. We thus have a long-time periodic evolution that combines fast Bloch-like oscillations with a slower evolution determined by the angular dependence of the force and the  backflow, which finally leads to a global strictly periodic evolution with a period much larger than the Bloch period $T$.

By varying the initial energy and velocity of the soliton, our study can be extended to qualitatively different adiabatic trajectories of this quantum pendulum. For low kinetic and potential ($\alpha \approx \pi/2$) energies, one recovers the behaviour of a classical pendulum, with an elliptical phase portrait trajectory and with $\tilde P<\pi$, corresponding to $w$ which is always 0 (see the ellipse in Fig.\,\ref{fig4}c). For large initial velocities, open trajectories are expected, corresponding to a wave packet making complete loops in the ring (see the unclosed line in Fig.\,\ref{fig4}c).

\begin{figure*}[t!!]
\includegraphics[width=0.35\textwidth]{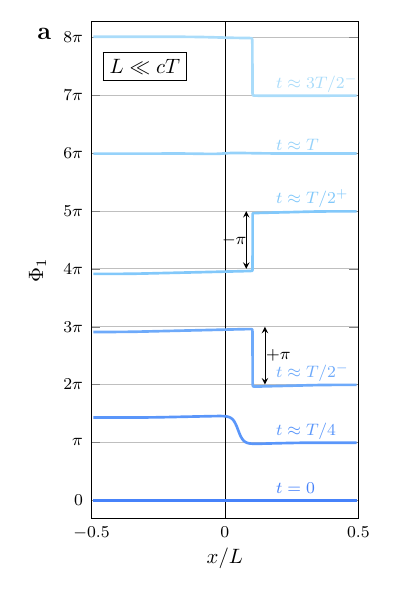}
\hskip-20pt
\includegraphics[width=0.35\textwidth]{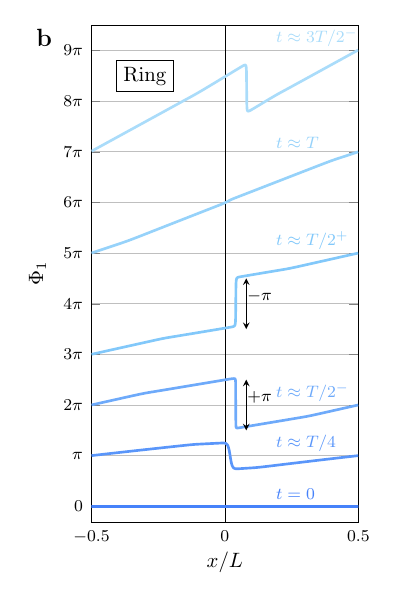}
\hskip-20pt
\includegraphics[width=0.35\textwidth]{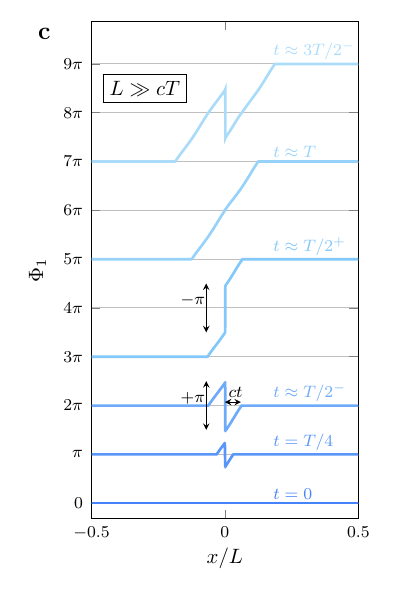}
\caption{\textbf{Comparison of the evolution of the bath phase for different geometries.} \textbf{a,}  Small system configuration, similar to the linear geometry explored in the main text. Boundary effects make the phase profile almost constant on each side of the wave packet. After one oscillation of the wavepacket,  the phase profile is uniform again leading to periodic Bloch oscillations.   \textbf{b,} Ring geometry. After one period, a phase winding of $2\pi$ is pumped into the bath. \textbf{c,} Long system. We observe a linear variation of the phase associated to the backflow over a distance $ct$. After one period a velocity field is present around the soliton.}
\label{fig:SMPhasePlot}
\end{figure*}

\subsection{Numerical simulations}
For weakly interacting Bose gases at zero temperature, the time evolution of the matter wave fields $\psi_1(x,t)$ and $\psi_2(x,t)$ associated with each component of the mixture is described by the two coupled NLSEs given in Eq.\,\eqref{eq: coupled_GPEs_magnetic}. The intraspecies interaction parameters are identical in our case and labelled $g$. The interspecies interaction parameter is labelled $g_{i}$.

We performed numerical simulations of the magnetic soliton dynamics given by these equations. The phase profiles shown in Fig.\,2 and Fig.\,3 of the main text are obtained from such simulations. The  parameters chosen for these simulations are similar to the typical experimental values. We used
$g=2.96\times10^{-39}$\,J.m, $g_{i}=1.0135 \,g$, $n_0=333\,\mum^{-1}$, $N_2=1300$ and $b'=1\,$G/m. The system sizes are $L=60\,\mum$ and $150\,\mum$ for the linear  and the ring geometry (with periodic boundary conditions), respectively. We show in Fig.\,\ref{fig:SMPhasePlot} the same data as in the main text, represented in a different way and compare it with the calculated phase profile in the case of a long linear system ($L\gg cT$), with strict boundary conditions. This situation, not experimentally studied in this work, is qualitatively different from the two cases studied in the main text. A rapid phase variation is still present around the wave packet position but the phase is not uniform on each side of the system. This is due to the finite speed of sound associated with the density mode which is not large enough to allow for the information to propagate over the full size of the system. In this case, the phase variation around the wave packet is compensated by a backflow associated with a linear spatial variation of $\Phi_1$ over a length $ct$. Thus, in a sufficiently large system, a current is present in the bath around the soliton and we expect a modification of the ``ideal'' BOs observed in a small system. 

\begin{figure}[!t]
\includegraphics[width=0.5\textwidth]{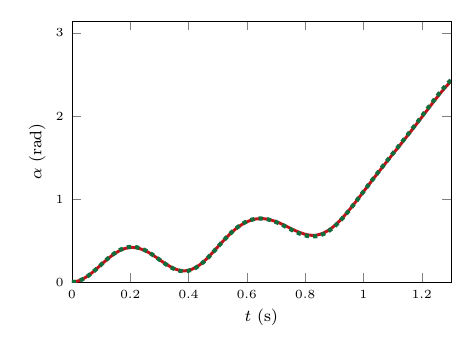}
\caption{\textbf{Benchmarking of the particle-like model in a ring geometry to the direct solving of the coupled NLSE.} Time evolution of the position of the center-of-mass of the wave packet when subjected to a constant force with $N_2 = 1100$, $b'=1\,$G/m, $n_0=330\,\mum^{-1}$. The red solid line is the the center-of-mass of the wave packet obtained by solving the coupled NLSE. The dashed-dotted green line is the prediction of the particle-like model.}
\label{fig:SMcomparisonGP}
\end{figure}

Numerical simulation can also be used to benchmark the particle-like model presented above. We show in Fig.\,\ref{fig:SMcomparisonGP} a quantitative comparison between the prediction of this particle-like model in the case of a ring geometry with the numerical solution of the coupled NLSE. We obtain an excellent agreement between the two models, which confirms, in the range of parameters studied in this work,  the validity of the description of the solitonic wave packet motion by the particle-like model in the adiabatic approximation.

%
%
%
%
%

\end{document}